\documentclass[prl,a4paper,twocolumn,showpacs]{revtex4-1} 
\usepackage{amsmath}
\usepackage{amsfonts}
\usepackage{amssymb}

\usepackage[english]{babel}
\usepackage[utf8]{inputenc}
\usepackage{ae}
\usepackage{latexsym}
\usepackage{bm}
\usepackage[dvips]{graphicx,color}
\usepackage{nonfloat}

\begin{document}

\title{From Inherent Structures Deformation to Elastic Heterogeneities}
\author{F. Léonforte}
\email{leonforte@theorie.physik.uni-goettingen.de}
\affiliation{                    
Institut für Theoretische Physik, Georg-August-Universität,\\
Friedrich-Hund-Platz 1, 37077 Göttingen, Germany
}

\date{\today}

\begin{abstract}
Using a well defined soft model glass in the framework of Molecular Dynamics simulations, the inherent structures are probed by means of a recently developed deformation protocol that aims to capture the Dynamical Heterogeneities (DH), as well as by the use of the \emph{isoconfigurational} ensemble. Comparisons of both methods are performed by extracting the corresponding inherent characteristic length scales as the temperature of the system is cooled down from the liquid to the glassy state. The obtained lengths grow and depict an identical trend as the system falls out-off equilibrium, and appear to converge to the characteristic length scale that characterizes the Elastic Heterogeneities (EH) of the materials in the very low temperature limit, which is deeply related to the properties of the glass. This provides a first evidence of a relationship between DH and EH.
\end{abstract}

\pacs{61.20.Lc,64.70.P-,64.70.qj}

\maketitle

\indent Nowadays, one of the challenging problem in the physics of glassy systems lies in bridging different concepts devised by recent theories of the glass transition~\cite{BookGlass2011,Ediger2012}. One aim is to improve and test them, but also to develop new mathematical tools and frameworks. Recently, two fundamental concepts seem to be of predominant interest. They deal with addressing the "structure-to-mobility" relashionship by accounting for the involved collective \emph{phenomenae} that appear as the glass transition is approached.\\
\indent To tackle this problem, most of the recent approaches develop a strategy in terms of monitoring a correlation length that has to grow as the glass forms. The nature, definition and pertinent choice of this length scale is still unclear~\cite{Kob2012}. From the "conventional" multi-point space-time correlation functions methods~\cite{Bouchaud2005,Biroli2006,Berthier2007,Berthier2005}, a "dynamical" correlation lenght emerges that only accounts for the heterogeneous dynamics but fails to address the structure-to-mobility relashionship. On the other hand, several theoretical frameworks compose with a "static" length scale that should characterize the growing emergence of spatially extent structures~\cite{Hocky2012,Biroli2008,Berthier2012,Karmakar2009}, but its connection with cooperative dynamical events is still a matter of debate. Furthermore, such a length scale is not confirmed experimentally and no consensus emerged regarding its definition. Recently, one method was proposed~\cite{Mosayebi2010,Mosayebi2012}, which aims to bridge the scale between inherent structure formalism of supercooled liquids~\cite{WC2009,WidmerCooper2004} and the elastic properties of amorphous solids~\cite{PreviousStudies} by obtaining and monitoring a characteristic length scale. The obtained length was proposed to reveal the trace of an associated critical length diverging at the glass transition, which in a theoretical framework derives from critical phenomena and Random First Order Transition theories (RFOT)~\cite{Kirkpatrick1989,Cavagna2009,Bouchaud2004,Tanaka2010}. In the following, this formalism is applied to a recently studied model glass~\cite{Leonforte2011b} for which "conventional" methods have been shown to not correctly account for the emergence of heterogeneities in the dynamics and elastic properties of the resulting glass.\\
\indent\textsc{Numerical model and Static length scale:} The systems under considerations have been well characterized in previous studies~\cite{PreviousStudies}. Briefly, they consist in slightly polydisperse two-dimensional Lennard-Jones glasses (2DLJ), composed of $N=10000$ particles at constant density, chosen in such a way that in the glassy state, the average pressure is close to zero, and that the glass mimics the behaviour of a soft glass. The critical temperature at which the system falls out-off equilibrium has been evaluated around $k_B T_c=0.24\epsilon$. The 2DLJ glass, already in a $k_B T\rightarrow 0$ state $X^0=\{\textbf{r}^0_i\}$, is then submitted to a macroscopic elongation applied on each particles $i$, such that  $\textbf{r}^0_i\rightarrow\textbf{r}^{0,d}_i\equiv\textbf{r}^0_i(1+\epsilon_{xx})$ directly followed by a minimization of the potential energy, which gives rise to the final state $\textbf{r}^{0,q}_i$. Following the notation of Ref.~\cite{Mosayebi2012}, this defines the DQ deformation, here in the athermal limit. The noisy part of the displacement field $(\textbf{r}^{0,q}-\textbf{r}^{0,d})$, i.e. \emph{non-affine} component, has been shown to depict collective motions of particles organized in large vortexes of characteristic size $\xi_{naff}\sim 30$ interatomic distances~\cite{PreviousStudies} that also determines the elastic heterogeneities (EH) length scale, and below which the classical continuum theory of elasticity is subject to strong limitations.\\
\indent\textsc{Inherent structures {\it via} deformation formalism (IQD):} A generalization of the above scheme to any initial $k_B T\neq 0$ state $X=\{\textbf{r}_i\}$ has been proposed in Ref.~\cite{Mosayebi2012}, which aims to capture dynamical "elastic" heterogeneities. It consists in first generating the DQ deformation, finally written as $X^{dq}\equiv\{\textbf{r}^{d+q}_i\}$, while another path is performed by first minimizing the initial state $X=\{\textbf{r}_i\}$, which gives the state $\textbf{r}^q_i$, and then deform it, leading to the path $X^{qd}\equiv\{\textbf{r}^{q+d}_i\}$. This second path is denoted as QD deformation. The comparison between the two DQ and QD paths $\textbf{d}_i\equiv\textbf{r}^{d+q}_i-\textbf{r}^{q+d}_i$ finally defines the \emph{$T$-non-affine} field. Both \emph{$T$-non-affine} and \emph{non-affine} displacement fields can be characterized by their magnitude. In the $k_B T\rightarrow 0$, the \emph{non-affine} field has been shown~\cite{PreviousStudies} to lie in the reversible deformation regime, as well as deformation rate independant for a range of values $10^{-5}\leq\epsilon_{xx}\leq 10^{-2}$. For the \emph{$T$-non-affine} field, the measure of its magnitude $d(k_B T,\epsilon_{xx})=\langle N^{-1}\sum_i \textbf{d}^2_i\rangle^{1/2}$ as a function of the temperature $k_B T$ is a way to quantify the same rate effects.

\begin{figure}[t]
	\includegraphics*[width=8cm]{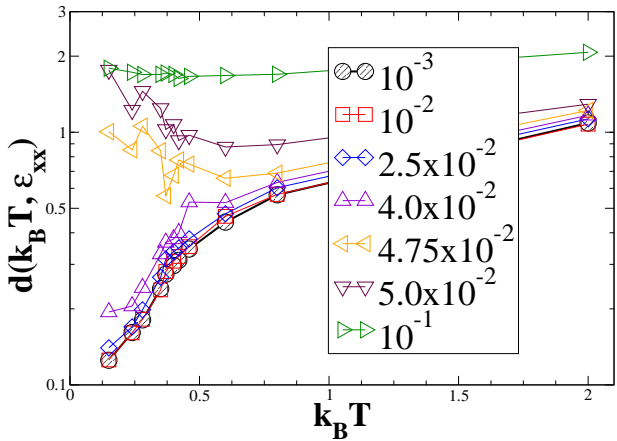}
	\caption{\label{fig.1}(Color online) Magnitude of the \emph{$T$-non-affine} at different deformation strength $\epsilon_{xx}$ and temperatures $k_B T$. The field becomes independant of the strength from which it was generated at each temperature, below $\epsilon_{xx}=2.5\times 10^{-2}$.}
\end{figure}
\indent Results are summarized in the Fig.\ref{fig.1}, which shows the magnitude of the \emph{$T$-non-affine} field at different deformation strengths and temperatures. It becomes independant of the applied deformation, i.e weak perturbation regime, for all the considered temperatures for values below $\epsilon_{xx}=2.5\times 10^{-2}$. As in the case of the $k_B T\rightarrow 0$ \emph{non-affine} field~\cite{Leonforte2011b}, one can interpret this weak perturbation regime to the reversible path of inherent structure deformations, namely that the energy landscape of local inherent structures is weakly deformed so that local structures can reversibly drop from one local minima to the closest neighbouring one, which is probably the same than the one before the perturbation is applied. For deformation rates $\epsilon_{xx} > 4.0\times 10^{-2}$, a minimum appears in the Fig.\ref{fig.1} that also marks the onset of rigidity of the material when the temperature is decreased. Indeed, when the material starts to fall out-off equilibrium, one expects that it becomes more "rigid" and "responsive" to the applied deformation. In the linear response regime, this should lead to a decrease of $d(k_B T,\epsilon_{xx})$. For stronger perturbations, more irreversible local paths are explored, which lead to an additional "plasticity" and drives the "noisy" \emph{$T$-non-affine} field to increase as $k_B T$ decreases. In the following, we only consider $\epsilon_{xx}<2.5\times 10^{-2}$ and therefore writes the \emph{T-non-affine} field like $\textbf{d}(k_B T)$.\\
\indent\textsc{Inherent structures {\it via} isoconfigurational ensemble formalism (ICE):} The inherent structures are explored in the framework of the \emph{isoconfigurational} ensemble (ICE)~\cite{WidmerCooper2004}, which probes the dynamical "mobile" heterogeneities of the material. Technically, we quench an initial state $X=\{\textbf{r}_i\}$ identical to the one used for the IQD, which leads to $X^{q}\equiv\{\textbf{r}^{q}_i\}$. From this state, one performs
$N_{r}\equiv 1000$ runs over a simulation period of $1.5\tau_{\alpha}(T)$, where $\tau_{\alpha}(T)$ is the $\alpha-$relaxation time of the model~\cite{Leonforte2011b}, while for each new runs momenta are randomly distributed from a Maxwell-Boltzmann distribution. Then, the squared displacement $\langle \Delta\mathbf{r}^2_i\rangle_{ic}$ characterizes the propensity of motion for a particle $i$.

\begin{figure}[t]
	\includegraphics*[width=8.5cm]{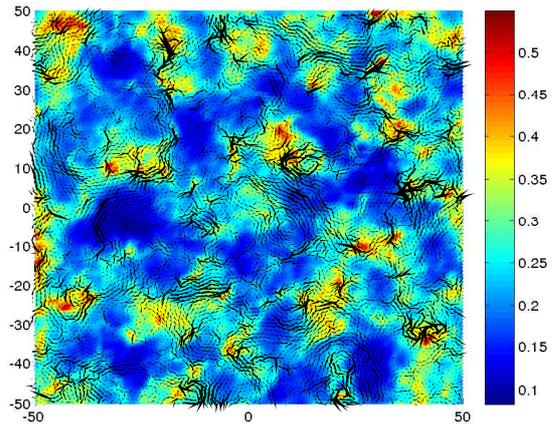}
	\caption{\label{fig.2}(Color online) Spatial distribution of propensities at $k_B T = 0.42\epsilon$ in ICE formalism compared to the corresponding \emph{T-non-affine} displacement field from IQD formalism. The same reference state $X=\{\textbf{r}_i\}$ is used. The colder the color, the lower the particle propensity, and vice versa for warmers.}
\end{figure}
\begin{figure}[t!]
	\includegraphics*[width=8.5cm]{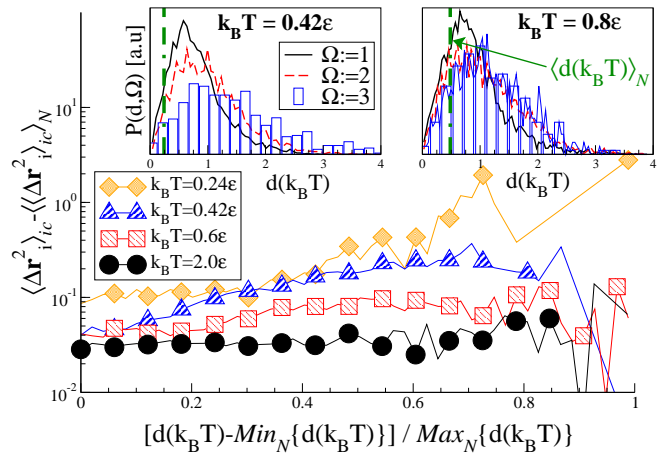}
	\caption{\label{fig.3}(Color online) {\it Main Panel:} histogram of the one-to-one correspondances between the norms of the \emph{T-non-affine} field for particles $i$ and their propensity. {\it Insets:} one-to-one distribution function $P\left(d(k_B T),\Omega\right)$ of the \emph{T-non-affine} field for particles $i$ belonging to propensity ensembles $\Omega$, and for $k_B T> k_B T_c$. The vertical dashed line corresponds to the average value $\langle d(k_B T)\rangle_{\textit{N}}$ (see text for details).}
\end{figure}

\indent\textsc{Definitions:}
\indent It is useful to separate
the propensity domains in three parts: $\Omega:\equiv 1\,\mid\,\lbrace\langle \Delta\mathbf{r}^2_i\rangle_{ic}\in [0.0,\overline{\Delta\mathbf{r}^2_i}\rbrace$, where $\overline{\Delta\mathbf{r}^2_i}$ is the mean propensity magnitude, $\Omega:\equiv 2\,\mid\,\lbrace\langle \Delta\mathbf{r}^2_i\rangle_{ic}\in [\overline{\Delta\mathbf{r}^2_i},w + \overline{\Delta\mathbf{r}^2_i}\rbrace$, where $w$ is the width of the distribution function of propensities that gives the mean $\overline{\Delta\mathbf{r}^2_i}$, and finally, $\Omega:\equiv 3\,\mid\,\lbrace\langle \Delta\mathbf{r}^2_i\rangle_{ic}\geq w + \overline{\Delta\mathbf{r}^2_i}\rbrace$. We therefore define the property $\mathcal{C}_{\Omega}$ as the condition that must fulfill a particle $i$ to be part of an ensemble $\Omega$ (bracketed part of the $\Omega$'s definitions), and we write the ensemble average for a quantity $X_i$ as $\langle X\rangle_{\Omega}=N^{-1}_{\Omega}\sum^N_{i=1}\delta\left(\mathcal{C}_{\Omega}-X_i\right)$, while $\langle X\rangle_N$ represents the normal average. Finally, in order to extract local quantities from both IQD and ICE formalisms, one defines the \emph{one-to-one} correspondance filter $\mathcal{F}_{\Omega}:=\lbrace (i,j)\in\Omega,\,\textrm{AND},\,\left(d_i(k_B T), d_j(k_B T)\right)\in\mathcal{A}_{\Omega}\rbrace$ such that $\mathcal{A}_{\Omega}:=\left[\langle d(k_B T)\rangle_{\Omega}-w_{\Omega};\langle d(k_B T)\rangle_{\Omega}+w_{\Omega}\right]$ where $w_{\Omega}=\sqrt{\textit{var}\lbrace P\left(d(k_B T),\Omega\right)\rbrace}$, and $P\left(d(k_B T),\Omega\right)$ is the \emph{one-to-one} distribution function like the ones depicted in the insets of the Fig.~\ref{fig.3}.\\
\indent\textsc{Results and discussion:}
\indent Fig.~\ref{fig.2} compares the propensity map at $k_B T=0.42\epsilon > k_B T_c$ to the \emph{T-non-affine} displacement field for the same initial configuration $X=\{\textbf{r}_i\}$ obtained using IQD. This highlights the correlations between the zones that follow the largest wavelength excitation (negligible $\textbf{d}(k_B T)$ or weakly perturbated inherent structures) and the highly probable immobile paths (lowest propensities). It also shows that to the softer zones in the energy landscape (large propensities) seem to correspond large fluctuating zones in terms of local inherent structure properties with larger $\textbf{d}(k_B T)$. This is rationalized in the Fig.~\ref{fig.3} where one plots the histograms of the \emph{one-to-one} correspondance between the norm $d_i(k_B T)$ and the corresponding $\langle \Delta\mathbf{r}^2_i\rangle_{ic}$ for a particle $i$. To get rid of the temperature dependance, the \emph{T-non-affine} fields are rescaled according to their minimal and maximal amplitude, respectively $\textit{Min}_N$ and $\textit{Max}_N$, which are calculated over the whole set of $N$ particles. As the temperature decreases, the distribution evolves from a flat one to an asymetric one, meaning that particles with high propensities also have a large $\textbf{d}(k_B T)$. This is enhanced as the systems falls out-off equilibrium and confirmed in the insets of the Fig.~\ref{fig.3}, where the \emph{one-to-one} distribution $P\left(d(k_B T),\Omega\right)$ is computed for particles that belong to an emsemble $\Omega$. For the lowest temperature $k_B T=0.42\epsilon$, it clearly shows that particles of high propensities, i.e. $\Omega:=3$, have a broad distribution of $d(k_B T)$ that also strongly departs from the average $\langle d(k_B T)\rangle_N$ by larger values. The opposite trend is observed for particles belonging to the low-level ensemble $\Omega:=1$. The same distributions are also plotted at an higher temperature $k_B T=0.8\epsilon$. One notes that they all look closer to the average $\langle d(k_B T)\rangle_N$ and that the \emph{one-to-one} correspondance distributions per ensemble $\Omega$ strongly overlap each others. At this temperature, the correspondances are mixed up and it is hard to correlate dynamically facilitated domains to the local structural and mechanical properties of the material.

\begin{figure}[t]
	\includegraphics*[width=8.5cm]{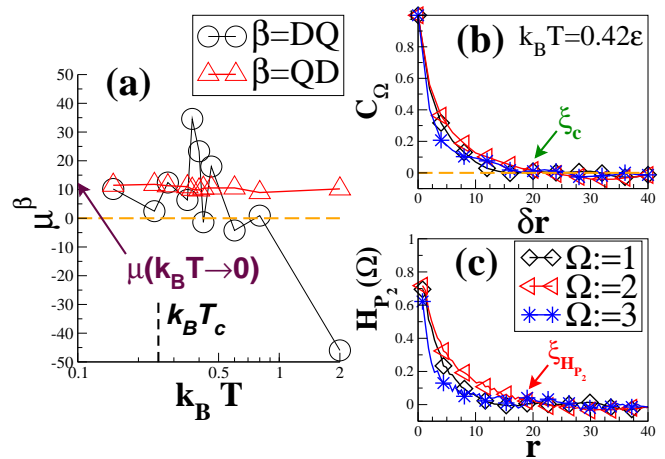}
	\caption{\label{fig.4}(Color online) {\bf (a)}: shear modulus obtained from the Hooke's law following the $\beta=\textrm{DQ or QD}$ branches, as a function of the temperature and for a deformation rate $\epsilon_{xx}=10^{-3}$. Its value in the athermal limit~\cite{PreviousStudies} is also given. {\bf (b)} and {\bf (c)}: obtention of the length scales $\xi_C$ and $\xi_{H_{P_2}}$ that capture the spatial correlations in the \emph{T-non-affine} field after use of the \emph{one-to-one} correspondance filter $\mathcal{F}_{\Omega}$ at $k_B T=0.42\epsilon$. The functions $C_{\Omega}$ and $H_{P_2}(\Omega)$ are described in the text.}
\end{figure}

\indent In the Fig.\ref{fig.4}(a), the IQD formalism is used in order to obtain the "elastic" coefficients through the Hooke's law~\cite{PreviousStudies}. This is done by estimating the difference in stresses, for the two DQ and QD branches, between the final $X^{dq}$ and $X^{qd}$ and their respective initial states. The two elastic Lamé coefficients derive from the relations $\lambda^{\beta}=\Delta\sigma^{\beta}_{yy}/\epsilon_{xx}$ and $\mu^{\beta}=\left(\Delta\sigma^{\beta}_{xx}-\Delta\sigma^{\beta}_{yy}\right)/2\epsilon_{xx}$, where $\beta\equiv\textrm{DQ or QD}$. The shear modulus $\mu^{\beta}\left(k_B T\right)$ are plotted in the Fig.\ref{fig.4}(a), for $\epsilon_{xx}=10^{-3}$ and for both $\beta$ protocols. One notes that the QD branch is insensitive to the temperature, mainly because the inherent structures are first probed and then weakly deformed, which can only lead to an average shear modulus that reflects the averaged intrinsic material properties of the glassy state. This is confirmed when one compares $\mu^{\textrm{QD}}\left(k_B T\right)$ to the athermal limit~\cite{PreviousStudies}. Conversely, the DQ branch first deforms and explore the pre-probed inherent structures and the possible neighbouring paths, and thus reflects the emergence of the intrinsic properties of the glass as the temperature is decreased. $\mu^{\textrm{DQ}}\left(k_B T\right)$ varies from negative (unstable) values to positive ones (resistance to deformation) when $k_B T$ decreases, and reaches $\mu^{\textrm{QD}}\left(k_B T\right)$ when $k_B T\approx k_B T_c$ at which the system starts to fall out-off equilibrium. From this temperature, inherent structures are formed and selected, and define the local and average properties of the material.

\indent Making use of the seeming complementarity of ICE and IQD, we then employ the filter $\mathcal{F}_{\Omega}$ in order to capture the distance $\mathbf{r}$ until which the fields $\mathbf{d}(k_B T)$ are correlated. The use of the filter ensures that the \emph{one-to-one} correspondance property is preserved. One first considers the correlation function $C_{\Omega}(\delta r)\equiv\langle\mathbf{d}(k_B T,\mathbf{r}+\mathbf{\delta r})\cdot\mathbf{d}(k_B T,\mathbf{r})\rangle_{\Omega}$ which has been shown to correctly capture the structural dominant correlations in noisy displacement fields~\cite{PreviousStudies,Chikkadi2012}. Such a function is plotted in the Fig.~\ref{fig.4}(b) for each ensemble $\Omega$, and from this are extracted the length $\xi_C$ above which the \emph{T-non-affine} field is not correlated anymore. Alternatively, the vortex-like structures of $\mathbf{d}(k_B T)$ and their location with specific propensity domains is quantified by computing the histograms of the second Legendre polynomial function $P_2=(1/2)\left(\langle\cos^2\theta\rangle - 3\right)$ where $\theta$ is the angle between two vectors, as a function of their distance from their origin, after applying the filter $\mathcal{F}_{\Omega}$. The characteristic length scale is therefore obtained by finding the distance $\mathbf{r}$ that obeys the property $\xi_{H_{P_2}}\equiv\left\lbrace \mathbf{r}\,\mid H_{P_2}(\mathbf{r},\Omega:=1)\cap H_{P_2}(\mathbf{r},\Omega:=2)\cap H_{P_2}(\mathbf{r},\Omega:=3)\right\rbrace$. The method is briefly illustrated in the Fig.\ref{fig.4}(c) for the system at $k_B T = 0.42\epsilon$ taken from the Fig.\ref{fig.2}. Independantly to the \emph{one-to-one} correspondance between ICE and IQD formalisms, we also characterize the structure of the \emph{T-non-affine} field using the coarse-grained correlation function~\cite{PreviousStudies,Mosayebi2012} defined as $B(b)=d(k_B T)^{-1}\left\langle\sum_{i=1}^N\left[N_i^{-1}\sum_{j=1}^{N_i} \textbf{d}_j(k_B T)\Theta_b(r_{ij})\right]\right\rangle$ where $\Theta_b(r_{ij}) = 1$ if $r_{ij} < b$, and $0$ otherwise, which is shown to be well approximated by an exponential decay of the form $B(b)\approx\exp{\left[-b/\xi_B\right]}$, and where $\xi_B$ characterizes the extent of correlations in the field.

\begin{figure}[t]
	\includegraphics*[width=8.5cm]{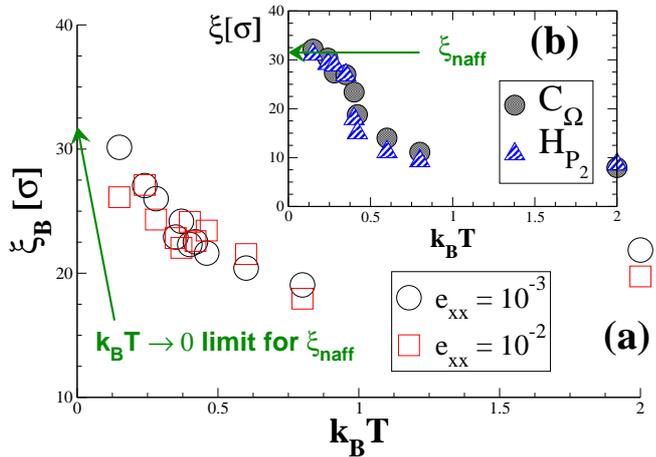}
	\caption{\label{fig.5}(Color online) Temperature dependance of the characteristic length scales of the \emph{T-non-affine} {\bf (a)}: $\xi_B$ from the exponential decay fit of $B(b)$, and for two deformation rates in the weak perturbation regime. {\bf (b)}: $\xi_C$ and $\xi_{H_{P_2}}$ after use of the \emph{one-to-one} correspondance filter $\mathcal{F}_{\Omega}$. For comparison is also given the $k_B T\rightarrow 0$ \emph{non-affine} characteristic length scale.}
\end{figure}
\indent Results for the temperature dependance of the derived length scales are compiled in Fig.~\ref{fig.5}. The coarse-graining correlation length $\xi_B$ is also plotted for two deformation rates below $\epsilon_{xx}=2.5\times 10^{-2}$. Its rate independence confirms the weak perturbation hypothesis, and its tendency to reach the athermal value $\xi_{naff}$ also validates the ability of the IQD formalism to capture the onset of cooperative effects that emerge upon cooling and lead to the final structural properties of the glass. In the Fig.~\ref{fig.5}(b) are plotted the length scales $\xi_C$ and $\xi_{H_{P_2}}$ obtained after use of the \emph{one-to-one} correspondance filter $\mathcal{F}_{\Omega}$. They characterize the spatial extent of domains of propensities motion of particles, and how these domains depict a specific inherent structure deformation ability, i.e. onset of mechanical response. A strong variation of both lengths therefore signals an increase of spatial heterogeneities in both an "elastic" and "mobility" sense. Such an increase upon cooling is indeed observed and, interestingly, the lengths appear to strongly increase around $k_B T\gtrapprox k_B T_c$, and then slowly converge, for $k_B T < k_B T_c$, to the athermal value $\xi_{naff}$ that characterizes the elastic response of the glass. Hence, the lengths also mark the onset of rigidity of the material, to which corresponds a selected distribution of elastic heterogeneities that has emerged from the distribution of mobile ones upon cooling.\\
\indent\textsc{Conclusion:} In this work, the inherent structures of a model glass were probed upon cooling, using a recently proposed deformation protocol in conjunction to the so-called \emph{isoconfigurational} ensemble. It has been shown that simultaneously using both methods provides a good framework that allows to estimate a characteristic length scale that can capture the dynamical heterogeneities (DH), correlate them to the emerging mechanical properties of the amorphous material, and finally recover the elastic heterogeneities (EH) of the glass.\\
\indent However, the obtained length scales do not seem to behave like critical ones~\cite{Mosayebi2010}, i.e. smooth temperature dependance, no divergence occurs, and no singularities appear around the Kauzmann temperature (extrapolated to $k_B T_K\simeq 0.162$), such that one can not actually discuss the nature of the critical behaviour of the liquid/glass phase transition. Additional investigations are needed to clarify these fundamental points.\\
\indent\textsc{Acknowledgements:} Financial support from grants of the German Science Foundation DFG-NSF Materials World Network (Mu1674/12) are acknowledge.

\end{document}